\documentclass[reprint,amsmath,amssymb,aps]{revtex4-1}
\pdfoutput=1
\usepackage{graphicx}
\usepackage{dcolumn}
\usepackage{bm}
\usepackage{natbib}
\begin{document}

\title{Subtle metastability of the layered magnetic topological insulator MnBi$_2$Te$_4$ from weak interactions}
\author{Jinliang Ning$^1$}
\author{Yanglin Zhu$^2$}
\author{Jamin Kidd$^1$}
\author{Yingdong Guan$^2$}
\author{Yu Wang$^2$}
\author{Zhiqiang Mao$^2$}
\author{Jianwei Sun$^1$}
	\email{Corresponding author: jsun@tulane.edu}
\affiliation{\mbox{$^1$Department of Physics and Engineering Physics, Tulane University, New Orleans, Louisiana 70118, USA} \\ $^2$Department of Physics, Penn State University, State College, Pennsylvania 16801, USA}

\date{\today}	

	\begin{abstract}
		The metastable layered compound MnBi$_2$Te$_4$ is the first experimentally realized intrinsic antiferromagnetic topological insulator, predicted to host the quantum anomalous Hall effect at high temperatures upon exfoliation to atomically thin layers. While its magnetic ordering and topological properties have generated intensive interest, the mechanism behind its metastability and the ideal crystal synthesis conditions have remained elusive. Here, using a combined first-principles-based approach that considers lattice, charge, and spin degrees of freedom, we investigate the metastability of MnBi$_2$Te$_4$ by calculating the Helmholtz free energy for the reaction Bi$_2$Te$_3$ + MnTe $\rightarrow$ MnBi$_2$Te$_4$. We identify a narrow temperature range (767 K to 873 K) in which the compound is stable with respect to the competing binary phases and successfully synthesize high-quality MnBi$_2$Te$_4$ single crystals using the Bi-Te flux method within this range. We also predict the various contributions to the total specific heat, which is consistent with our experimental measurements. Our findings indicate that the degrees of freedom responsible for the van der Waals interaction, magnetic coupling, and nontrivial band topology in layered materials not only enable emergent phenomena but also determine thermodynamic stability. This conclusion lays the foundation for future computational material synthesis of novel layered systems. 
	\end{abstract}

	\maketitle
	
	\textit{Introduction}. MnBi$_2$Te$_4$ is an intrinsic antiferromagnetic topological insulator (TI) currently under intensive study  \cite{SCZ}, in which the interplay between magnetism and topology is expected to cause the quantum anomalous Hall (QAH) effect upon exfoliation to atomically thin layers \cite{SCZ,QAHE}. While bulk MnBi$_2$Te$_4$ is characterized by a $\mathbb{Z}_2$ invariant protected by the combination of time-reversal and half-lattice translational symmetry, the QAH insulator phase is enabled by a nonzero Chern number ($\mathbb{Z}$ invariant) in odd-layer thin films \cite{SCZ,tenfold}. MnBi$_2$Te$_4$ is formed by intercalating magnetic MnTe layers, wherein the open $d$-shell Mn ions form localized magnetic moments, into the quintuple layers of Bi$_2$Te$_3$, which is topologically protected due to spin-orbit coupling (SOC)-induced band inversions \cite{FKM,bansil}. Hence, as-synthesized MnBi$_2$Te$_4$ is also a layered material bound by the van der Waals (vdW) interaction.  
	
	The experimental synthesis of large, pure, high-quality MnBi$_2$Te$_4$ single crystals suitable for magnetotransport measurements is notoriously difficult. The first reported polycrystalline synthesis of MnBi$_2$Te$_4$ was achieved via congruent melting of stoichiometric melts of raw Bi, Mn, and Te, followed by rapid quenching, then annealing at 808 K \cite{poly}. Additionally, by using the binary phases as precursors, single crystals of MnBi$_2$Te$_4$, stable in a narrow temperature range below 873 K, have been synthesized from a solid-state reaction of Bi$_2$Te$_3$ and $\alpha$-MnTe followed by water-quenching and annealing at 838.15 K for 10 days \cite{single}. However, the obtained samples likely suffered from antisite defects resulting in nonstoichiometry \cite{single}, which could prevent the topological protection of the material due to magnetic disorder. While the powder samples were found to be metastable at room temperature \cite{poly}, bulk single crystals could be cooled to low temperatures without decomposition \cite{single}. More recently, large single crystals of MnBi$_2$Te$_4$ have been obtained using a Bi-Te flux with molar ratio Mn:Bi:Te = 1:10:16 \cite{flux}. The self-flux and vertical Bridgman  methods have also been used, both requiring extremely precise control of temperatures \cite{QAHE,bridgman}. The above experiments all confirmed that MnBi$_2$Te$_4$ is a metastable phase that can only be synthesized within a narrow temperature range below 873 K. \cite{poly, single, flux, bridgman}. Without a fundamental understanding of its metastability, the ideal growth conditions for pure, large crystal synthesis will remain elusive, ultimately preventing further experimental study of its unique topological and magnetotransport properties. Furthermore, although there have been a significant number of materials predicted to host similarly interesting topological properties, there are only handful of experimental realizations of such predictions \cite{catalogue1,catalogue2,comprehensive,fantasy}. Hence, the understanding of the metastability of MnBi$_2$Te$_4$ is also highly desirable for the synthesis of other candidate layered topological materials.
	
	\begin{figure}[t]
	\centering
	\includegraphics[scale=0.5]{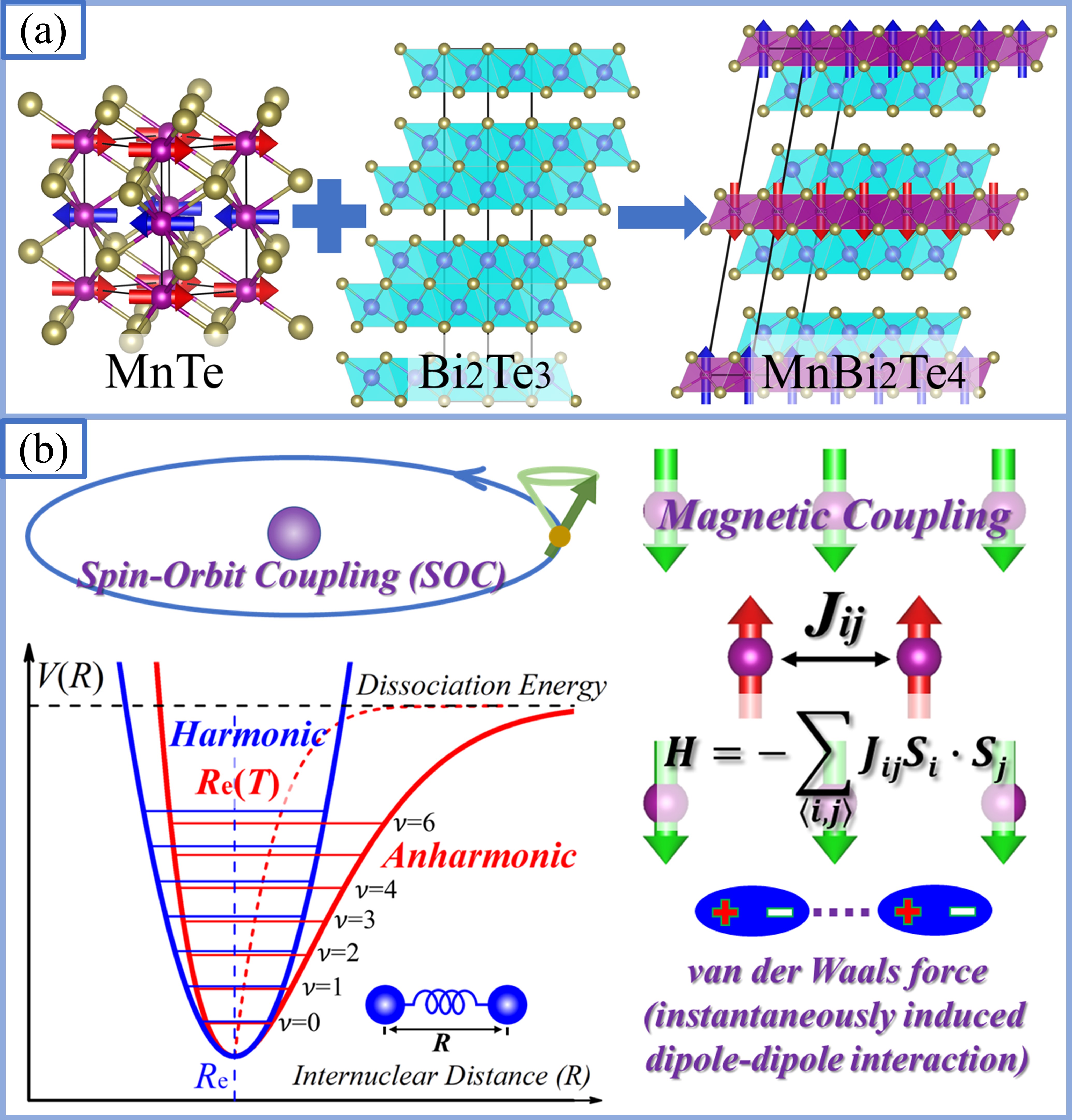}
	\caption{(a) Schematic of the synthesis reaction of the target ternary layered magnetic material MnBi$_2$Te$_4$ from two competing binary phases: Bi$_2$Te$_3$ + MnTe $\rightarrow$ MnBi$_2$Te$_4$. Note the layered structures of Bi$_2$Te$_3$ and MnBi$_2$Te$_4$, and the localized magnetic moments (illustrated by blue and red arrows on Mn) in MnTe and MnBi$_2$Te$_4$. (b) Schematic of the weak interactions considered here, found to be critical in determining the metastability of MnBi$_2$Te$_4$. Top left: SOC; top right: magnetic coupling interaction; bottom left: anharmonic nature of lattice vibration in comparison with the harmonic approximation; bottom right: vdW interaction.}
	\end{figure}

	Thermodynamic stability of materials can be predicted from chemical reaction free energy based on first-principles density functional theory (DFT) calculations. Nevertheless, for complex layered magnetic quantum materials like MnBi$_2$Te$_4$, such predictions are challenging, due to the increasing importance of various weak interactions (including SOC, magnetic coupling, vibrational anharmonicity, and vdW interactions), most of which are also responsible for their emergent properties, as illustrated in Figure 1. Typically, these materials have different kinds of chemical bonds ranging from strong in-plane covalent bonds to weak interlayer vdW interactions, with interaction strengths across almost 3 orders of magnitude (from {\raise.17ex\hbox{$\scriptstyle\sim$}}1 eV for strong bonds to {\raise.17ex\hbox{$\scriptstyle\sim$}}1 meV for vdW interactions). Both strong chemical bonds and weak interactions are equally important for the thermodynamic stability predictions, demanding a single density functional approximation for simultaneously accurate descriptions. More importantly, the electronic, magnetic, and vibrational thermal excitation energies of these materials at finite temperature can be up to several tens of meV; although this scale is typically negligible for stability predictions of bulk solids \cite{solids}, it is comparable to the thermal energy required to stabilize MnBi$_2$Te$_4$. Since DFT is a zero-temperature ground state electronic structure method, post-DFT models have to be used to describe such excitations. Therefore, all these interactions have to be considered and described accurately for the thermodynamic stability of MnBi$_2$Te$_4$, either at the DFT level or by DFT-based models.
	
	\begin{figure}[t]
		\centering
		\includegraphics[scale=0.5]{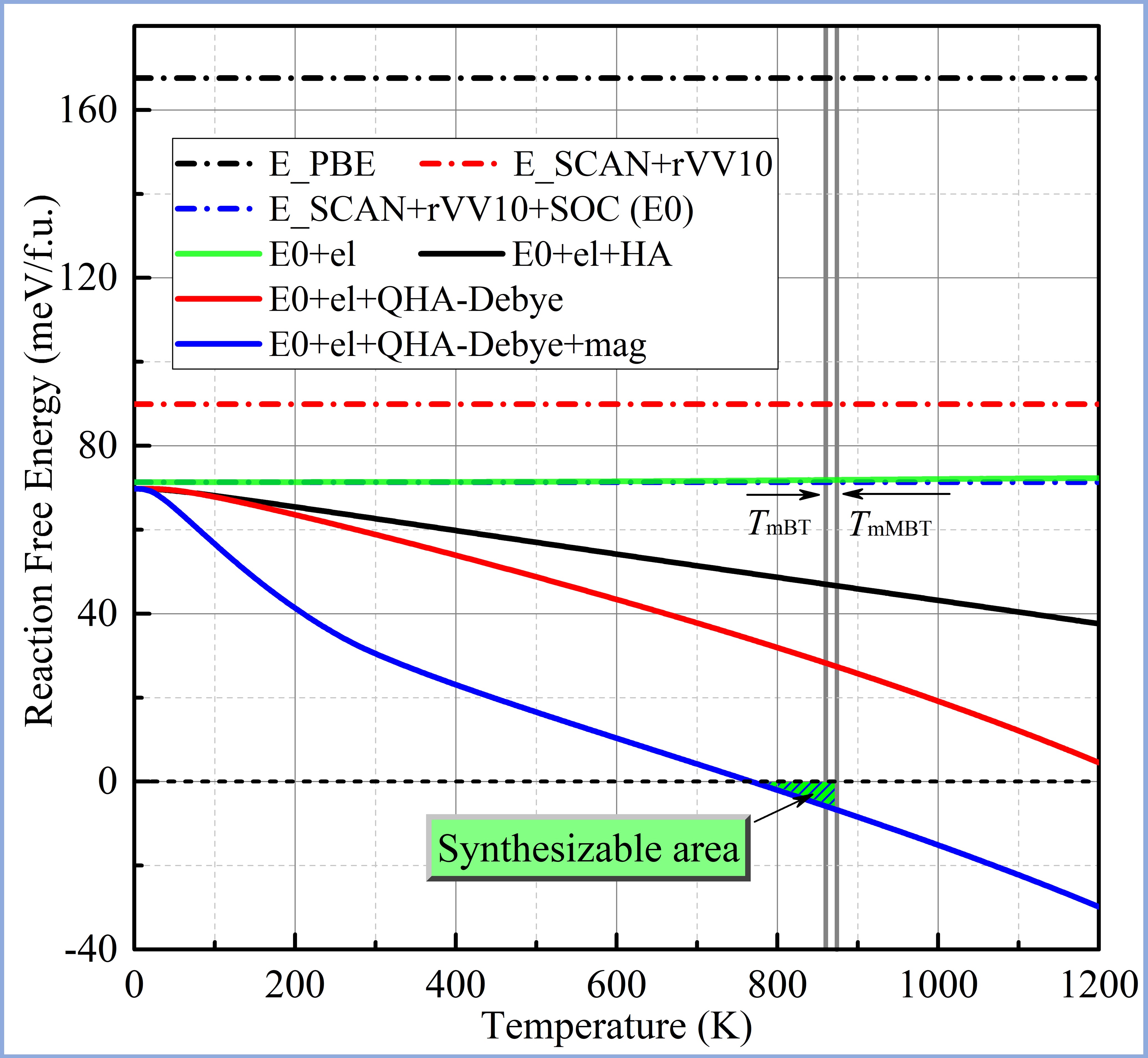}
		\caption{Reaction free energy based on Bi$_2$Te$_3$ + MnTe $\rightarrow$ MnBi$_2$Te$_4$, calculated at different levels of theory, including the 0 K DFT reaction energy from PBE (E$\_$PBE), SCAN+rVV10 (E$\_$SCAN+rVV10), and SCAN+rVV10 with SOC (E$\_$SCAN+rVV10+SOC, denoted as E0); the electronic excitation (el), lattice vibrational (harmonic approximation, HA, and quasi-harmonic approximation with Debye model, QHA-Debye), and magnetic (mag) contributions. Experimental melting points of Bi$_2$Te$_3$ and MnBi$_2$Te$_4$ are indicated by two vertical grey lines.}
	\end{figure}
	
	Here, using a set of first-principles-based approaches that consider lattice, charge, and spin degrees of freedom to take into account the various weak interactions, we calculate the temperature-dependent reaction free energy of MnBi$_2$Te$_4$ based on the reaction MnTe + Bi$_2$Te$_3$ $\rightarrow$ MnBi$_2$Te$_4$, which identifies a narrow temperature range for the metastability of MnBi$_2$Te$_4$ consistent with experiments. These approaches are then validated by comparing the predicted specific heat capacities of the relevant compounds to experimental results.  
	
	After all interactions are taken into account, Figure 2 shows that MnBi$_2$Te$_4$ is unstable with respect to its competing binary phases (MnTe and Bi$_2$Te$_3$) for low temperatures, and becomes stable above 767 K, indicated by the calculated reaction free energy (solid blue line).  However, since the melting points are {\raise.17ex\hbox{$\scriptstyle\sim$}}853 K and {\raise.17ex\hbox{$\scriptstyle\sim$}}873 K for Bi$_2$Te$_3$ and MnBi$_2$Te$_4$, respectively \cite{single}, the quenching required for single crystal synthesis of MnBi$_2$Te$_4$ without the minor Bi$_2$Te$_3$ phase must occur between {\raise.17ex\hbox{$\scriptstyle\sim$}}853 K and {\raise.17ex\hbox{$\scriptstyle\sim$}}873 K. This narrow temperature window, in addition to the fact that MnBi$_2$Te$_4$ is only more stable than the competing binaries by about 6.8 meV/f.u. in this window, explains why previous experimental attempts at synthesizing large single crystals have been difficult. 
	
	Below, we analyze the importance of various weak interactions to the metastability of MnBi$_2$Te$_4$. We find that the vdW interaction, SOC, magnetic ordering, and lattice vibrations (phonons) play crucial roles in determining the thermodynamic stability of MnBi$_2$Te$_4$ and tend to stabilize MnBi$_2$Te$_4$ with respect to the competing binary phases as temperature increases, while the electronic thermal excitation is negligible.

	\textit{Van der Waals Interaction}. It is well known that layered materials are bound by vdW interactions between layers, while the popular Perdew-Burke-Ernzerhof (PBE) density functional \cite{PBE} misses most vdW interactions and barely binds layered materials \cite{PBElayer,SCAN+}. Meanwhile, it has been shown that the strongly-constrained and appropriately-normed (SCAN) density functional \cite{SCAN1,SCAN2} combined with the revised Vydrov-Van Voorhis (rVV10) nonlocal correlation density functional \cite{rVV10} is effective in accurately describing structural and energetic properties of layered materials \cite{SCAN+}. As expected, Figure 2 shows that PBE destabilizes  MnBi$_2$Te$_4$ with respect to the competing binaries by about 80 meV/f.u. in comparison with SCAN+rVV10 for the 0 K ground state energies. Since 80 meV/f.u. is significantly larger than the finite temperature contributions to the reaction free energy from the electronic, magnetic, and vibrational degrees of freedom, PBE will falsely predict that MnBi$_2$Te$_4$ cannot be synthesized below its melting temperature, even in a metastable phase. In addition to the vdW interaction, it is likely that SCAN+rVV10 more accurately describes the $d$ electrons of Mn (and hence the magnetic properties of MnTe and MnBi$_2$Te$_4$) when compared to PBE, due to the reduced self-interaction errors, a result demonstrated by previous studies on many other transition metal compounds \cite{SCAN1,SCAN2,Sun2013,MnO2,TMM,Furness2018}.

	\textit{Spin-orbit Coupling}. It is usually assumed that SOC is a purely atomic effect and canceled out in a reaction free energy calculation \cite{solids}. Figure 2 shows that the inclusion of SOC stabilizes MnBi$_2$Te$_4$ with respect to MnTe and Bi$_2$Te$_3$ by about 10 meV, which is negligible for typical solid reactions but is critical for the metastability of MnBi$_2$Te$_4$. It is well known that SOC becomes important for heavy atoms, like Bi and Te, and is responsible for the topological properties of Bi$_2$Te$_3$ and MnBi$_2$Te$_4$, while the open $d$-shell of Mn provides localized magnetic moments for MnTe and MnBi$_2$Te$_4$. The synergy of SOC and AFM-ordered localized magnetic moments results in MnBi$_2$Te$_4$ being an intrinsic magnetic TI \cite{MTI}. Our results further suggest that the presence of the magnetic moments enhances the SOC to lower the total energies more for MnBi$_2$Te$_4$ and MnTe than for Bi$_2$Te$_3$ (see supplementary materials, Figure S1). 

	\textit{Electronic Contribution}. Figure 2 shows that the electronic contribution to the reaction free energy is negligible. This is to be expected, since the relevant compounds are semiconductors with band gaps no less than 0.1 eV (see supplementary materials, Figure S2), an energy level much larger than the electronic thermal excitations for the considered temperature range. 
	
	\textit{Vibrational Contribution (Harmonic Approximation and Anharmonicity)}. Figure 2 shows that lattice vibrations under the harmonic approximation (HA) destabilize the competing binaries more than the ternary phase. This is because the MnTe binary is much stiffer than the other two compounds, as illustrated by their equations of state (see supplementary materials, Figure S3), and thus has much fewer phonon modes at low frequencies in the computed phonon density of states (see supplementary materials, Figure S4). We also found that the lattice anharmonicity modeled by the Quasi-harmonic Debye (QHA-Debye) model \cite{gibbs2.1,gibbs2.2} and the Slater approximation \cite{gibbs2.2} to the Debye temperature $\Theta_D$ stabilizes MnBi$_2$Te$_4$ with respect to MnTe and Bi$_2$Te$_3$. This is probably due to the fact that MnTe is stiffer in its equation of state and experiences less anharmonicity in the temperature range considered here than the other two compounds. The anharmonicity has an effect of softening the phonon modes and therefore likely reduces the phonon frequencies of MnBi$_2$Te$_4$ and Bi$_2$Te$_3$ more than MnTe, favoring thermal population. This effectively stabilizes MnBi$_2$Te$_4$ with respect to MnTe and Bi$_2$Te$_3$ as temperature increases. 

	\textit{Magnetic Contribution}. MnBi$_2$Te$_4$ has roughly the same Mn magnetic moment as MnTe, {\raise.17ex\hbox{$\scriptstyle\sim$}}4.3 $\mu_B$ \cite{MnMag}, while the exchange coupling strength $J_{ij}$ is much weaker (see supplementary materials, Figures S5 and S6), likely due to the layered structure. Similar to the lattice vibration effect, the stronger magnetic coupling should result in higher frequency magnons that destabilize MnTe more than MnBi$_2$Te$_4$ as temperature increases, as illustrated in Figure 2.
	
	\begin{figure}[!ht]
		\centering
		\includegraphics[scale=0.48]{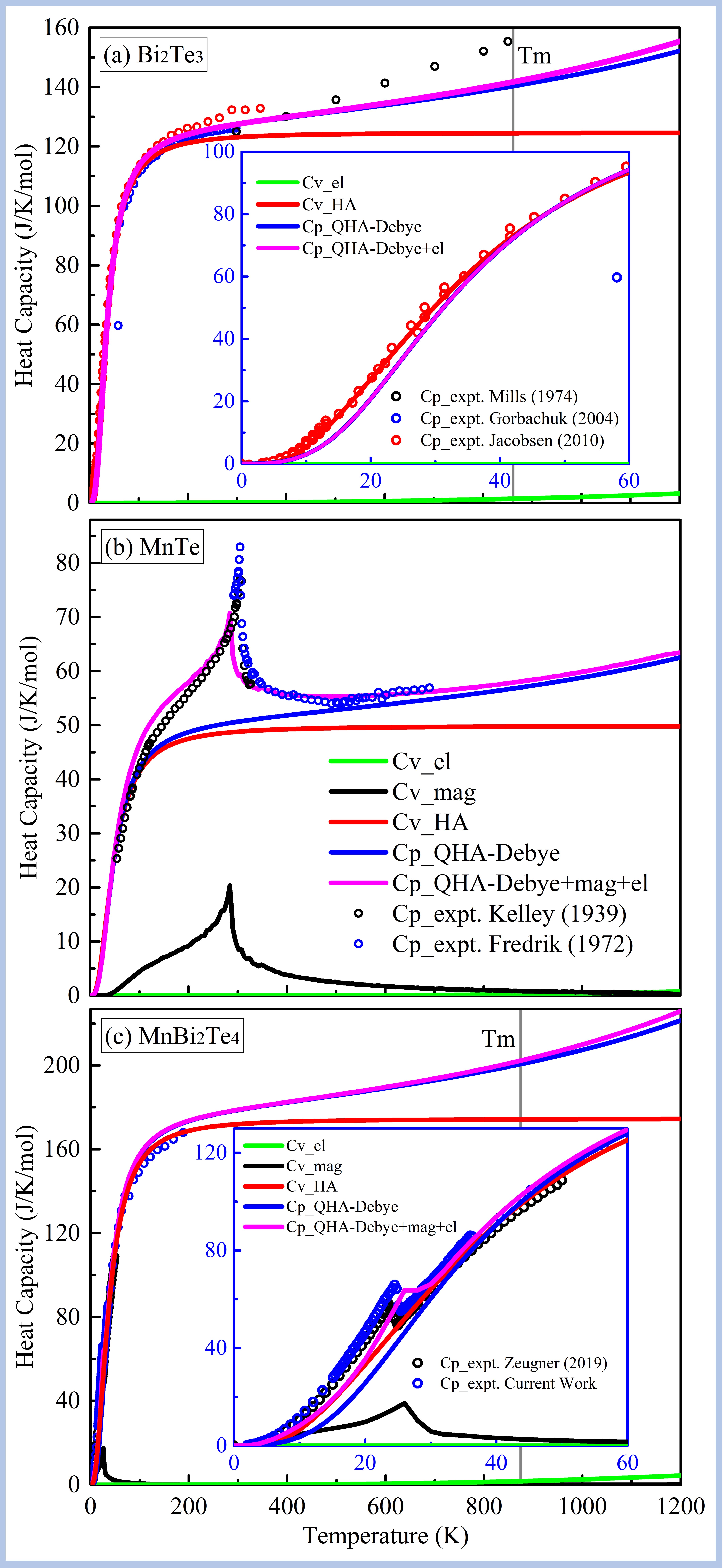}
		\caption{Calculated heat capacity from electronic, magnetic, and vibrational contributions, compared with available experimental data, for (a) Bi$_2$Te$_3$ \cite{mills, gorb,jacobsen}, (b) MnTe \cite{kelley,fredrik}, and (c) MnBi$_2$Te$_4$ \cite{single}. Cv$\_$el, Cv$\_$mag and Cv$\_$HA are contributions from electronic, magnetic, and vibrational (HA) excitations calculated within the constant volume approximation at 0 K equilibrium volume. Cp$\_$QHA-Debye is the constant pressure specific heat, involving the thermal expansion contribution.}
	\end{figure}
	
	\textit{Specific Heat}. Next, we turn to the calculated heat capacity at constant pressure ($C_p$) of the ternary phase MnBi$_2$Te$_4$ and its competing binary phases Bi$_2$Te$_3$ and MnTe, as shown in Figure 3. We choose to focus on $C_p$ because it is closely related to the free energy and is the key quantity obtained from calorimetric measurements that can be used to validate our calculations. Here, we neglect the thermal expansion effect on the electronic and magnetic contributions to $C_p$ and thus use $C_v$ for these two contributions instead. A complete description of the calculation is included in the computational methods section. 

	When compared to the most recent experimental data for each phase, our calculations are accurate below 300 K. Since heat capacity is closely related to the free energy, this result reinforces the reliability of our initial stability estimation for MnBi$_2$Te$_4$. Notable discrepancies between our calculations and experimental results for Bi$_2$Te$_3$ (see Figure 3a) occur above {\raise.17ex\hbox{$\scriptstyle\sim$}}400 K. This is likely due to the formation of point defects at high temperatures, such as vacancies and antisite defects, which has been consistently observed in experiment \cite{single}. While these defects affect $C_p$, entropy, and free energy for Bi$_2$Te$_3$ at high temperatures, similar formation of defects can be expected for MnBi$_2$Te$_4$ \cite{single}. Thus, the effects of defects in the total reaction free energy can mostly be canceled out. For both MnTe (T$_N$ $\approx$ 300 K) and MnBi$_2$Te$_4$ (T$_N$ $\approx$ 25 K), we predict well-defined peaks in $C_p$ near the respective antiferromagnetic phase transition temperatures, consistent with experimental data. 

	Specific heat measurements were also taken up to {\raise.17ex\hbox{$\scriptstyle\sim$}}200 K on a single crystal sample of MnBi$_2$Te$_4$ we synthesized using the Bi-Te flux method. When compared to our calculated results, the low-temperature $C_p$ is quite accurate (see inset, Figure 3c). At intermediate temperatures ({\raise.17ex\hbox{$\scriptstyle\sim$}}423 K to {\raise.17ex\hbox{$\scriptstyle\sim$}}767 K), thermal vibration is strong enough to excite MnBi$_2$Te$_4$ out of the metastable state and separate it into the binary phases \cite{poly}, which makes $C_p$ measurements nearly impossible. 
	
	
	\textit{Summary}. Based on first-principles thermodynamics, we have analyzed the metastability of the recent antiferromagnetic TI candidate MnBi$_2$Te$_4$ and have successfully synthesized high-quality single crystals for specific heat measurements. Our DFT-based approach yields specific heat results in good agreement with experimental data, validating our stability estimation of MnBi$_2$Te$_4$ as a function of temperature. We confirm that MnBi$_2$Te$_4$ is a metastable phase which can only be synthesized in a short range of temperatures (767 K to 873 K) with a small reaction free energy less than {\raise.17ex\hbox{$\scriptstyle\sim$}}6.8 meV/f.u., explaining the previous difficulties confronted in experiment. We demonstrate that fundamental weak interactions, including SOC, vdW, magnetic coupling, and lattice vibrations all contribute to this subtle high-temperature metastability. This finding potentially facilitates future computational discoveries of novel stable or metastable 2D magnetic topological materials.
	 \\[12pt]
	
	\textit{Experimental methods}. Single crystals of MnBi$_2$Te$_4$ were synthesized using the Bi-Te flux method \cite{flux}. The starting materials of Mn, Bi, and Te powder were mixed with a molar ratio of 1:10:16 and loaded into an Al$_2$O$_3$ crucible, then sealed in a quartz tube under high vacuum. The mixture was heated to 1173.15 K in a muffle furnace and held there for 24 hours to allow homogeneous melting, then slowly cooled down to 863.15 K at a rate of 3 K/h. After removing the excess Bi$_2$Te$_3$ flux through centrifuging, plate-like single crystals were obtained. Their crystal structure was confirmed by X-ray diffraction measurements, and their chemical composition was examined using energy dispersive X-ray spectroscopy. The specific heat of MnBi$_2$Te$_4$ up to 200 K was measured with an adiabatic relaxation technique using a commercial Physical Property Measurement System (PPMS, Quantum Design). This data is plotted in Figure 3(c).
	
	\textit{Computational methods}. In this work, we carry out DFT \cite{KSDFT} calculations using the Vienna Ab-initio Simulation Package (VASP) \cite{VASP} with the projector-augmented wave (PAW) method \cite{PAW1,PAW2}. The recently developed strongly-constrained and appropriately-normed (SCAN) meta-GGA \cite{SCAN1,SCAN2} is used for its superior performance in description of different chemical bonds and transition metal compounds \cite{SCAN1,SCAN2,Sun2013,MnO2,TMM,Furness2018}. A long range vdW correction is combined with SCAN through the rVV10 nonlocal correlation \cite{rVV10}, a revised form of VV10, the Vydrov-Van Voorhis non-local correlation functional \cite{VV10}. The PAW method is employed to treat the core ion-electron interaction. An energy cutoff of 520 eV is used to truncate the plane wave basis. We use $\Gamma$-centered meshes with a spacing threshold of 0.15 \AA$^{-1}$ for K-space sampling. Geometries of the three compounds were allowed to relax until the maximum ionic forces were below a threshold of 1 meV \AA$^{-1}$. We use the Phonopy code \cite{phonopy} to calculate the key information of phonon frequency $\omega_q$ (where $q$ is the wave vector), from harmonic force constants calculated by VASP within the DFPT method. The Perdew-Burke-Ernzerhof GGA \cite{PBE} is used for this purpose due to its smoother potential energy surfaces for generating forces \cite{smoother}.
	
	The key quantity for thermodynamic stability predictions of solids is the Helmholtz free energy, defined as
	$$F\left(V,T\right)=E0\left(V\right)+F^{el}(V,T)+F^{vib}(V,T)+F^{mag}(V,T)_{\textstyle.}$$
	Here, the $PV$ contribution has been ignored, and the adiabatic approximation has been used, which decouples lattice, charge, and spin degrees of freedom for thermal excitations. $P$ is the pressure, $T$ the temperature, $V$ the volume, and $E0\left(V\right)$ the total electronic energy at zero temperature, which can be directly calculated using different density functional approximations. $F^{el}\left(V,T\right)$,  $F^{vib}\left(V,T\right)$, and $F^{mag}\left(V,T\right)$ are the contributions to the free energy at finite temperature from electronic, lattice, and magnetic degrees of freedom, respectively, which can be modeled based on DFT results. In this study, $E0\left(V\right)$ was calculated using SCAN+rVV10.
	
	Based on the electronic density of states obtained from DFT calculations (see supplementary materials, Figure S2), the electronic contribution to the free energy, $F^{el}(V,T)$, can be determined by the finite temperature method \cite{finite} according to Fermi-Dirac distribution, following the fixed density of states approximation \cite{fixed}.
	
	The lattice vibrational contribution $F^{vib}\left(V,T\right)$ can be expressed as
	$$\frac{1}{2}\sum_{q,s}{\hbar\omega(q,s)}+k_BT \sum_{q,s} \ln \left[ 1 - \exp \left(\frac{\hbar\omega(q,s)}{k_BT}\right)\right]_{\textstyle,}$$
	where $\omega\left(q,s\right)$ is the phonon frequency associated with wave vector $q$ and band index $s$, and $k_B$ is the Boltzmann constant. $\omega\left(q,s\right)$ typically depends on $V$ and $T$ due to the anharmonicity of the lattice potential, and it can be calculated or modeled based on DFT at different levels. The HA assumes that the lattice sees a harmonic potential at the equilibrium volume (see Figure 1), and $\omega(q,s)$ is calculated based on the frozen-phonon approach, which computes the harmonic force constants from DFT calculations via finite atomic displacements \cite{phonopy}. The quasi-harmonic approximation (QHA) method \cite{phonopy,QHA} takes a step further to consider the volume dependence for the phonon anharmonicity, while the temperature is assumed to indirectly affect phonon vibrational frequencies through thermal expansion. Typically, the phonon spectra of about ten or more volumes are usually required for a DFT-based QHA simulation, and such calculations are always time-consuming. Here, $F^{vib}\left(V,T\right)$ are instead approximated using the phonon density of states (see supplementary materials, Figure S4) by the Debye model. The Debye temperature is then estimated based on the equation of state from DFT calculations to account for the anharmonicity \cite{gibbs2.1, gibbs2.2}.
	
	
	We fit the static $E(V)$ curve of each phase (see supplementary materials, Figure S3) to the Vinet equation of state \cite{gibbs2.1,vinet1,vinet2,vinet3,vinet4}. The total lattice vibrational contribution $F^{vib}(V,T)$ is treated by combining the HA (as implemented in Phonopy \cite{phonopy}) and the QHA-Debye model (as implemented in Gibbs2 \cite{gibbs2.1,gibbs2.2}) via a scaling factor (a function of the Poisson ratio) to recover the behavior of the HA at low temperature, similar to previous studies \cite{scaling1,scaling2,scaling3}. We chose not to use the conventional QHA method \cite{qha2}, not only because it is much more computationally expensive, but also because it requires more careful validation for layered systems (like GeSe \cite{GeSe}) and even some conventional materials (like Si \cite{Si1,Si2}), in addition to the ubiquitous imaginary frequency problems \cite{GeSe,imaginary}.
	
	An often-employed approach to capture $F^{mag}(V,T)$ of systems with localized magnetic moments starts with the effective Heisenberg Hamiltonian $$\hat{H}=-\frac{1}{2}\sum_{i,j}J_{ij}\hat{S}_i\hat{S}_j{}_{\textstyle,}$$ mapped from DFT calculations, with first-principles-derived exchange coefficients $J_{ij}$ (see supplementary materials, Figures S5 and S6) that mediate the magnetic exchange between spins $\hat{S}$ localized at lattice sites $i$ and $j$. $F^{mag}(V,T)$ can then be calculated by the eigenenergies of $\hat{H}$, similar to the previous expression for $F^{vib}(V,T)$. However, for most realistic systems, an exact analytical solution is not known. Here, $F^{mag}(V,T)$ corresponding to $\hat{H}$ is calculated using a rescaled Monte Carlo (rMC) method \cite{rmc}, which maps the classical Monte Carlo (CMC)-obtained thermodynamic quantities to those obtained by quantum Monte Carlo (QMC), scaled by a factor dependent on the spin quantum number $S$. Although the QMC solution to $F^{mag}(V,T)$ is desired, it has a limited applicability due to the so-called (negative) sign problem \cite{sign1,sign2}. CMC is known to be useful and reliable for magnetic critical temperature predictions and acceptable for high temperature (above critical temperature) heat capacity predictions. At 0 K, however, CMC gives a finite heat capacity. This is incorrect, since the zero-temperature heat capacity contributed by magnons should be zero according to quantum statistics of the magnon excitations. The use of rMC corrects the low temperature part effectively. 	
	
	Explicit coupling terms, e.g., phonon-magnon, phonon-electron, and higher order phonon-phonon interactions are all assumed to be small and neglected. Formation of defects at high temperatures (such as vacancies and antisite defects \cite{single}), especially for Bi$_2$Te$_3$ and MnBi$_2$Te$_4$, may also contribute to the $C_p$ and the free energy. This defect factor, however, is expected to mostly cancel between the two phases and is possibly negligible to the total free energy. Therefore, this factor is not considered in our calculation.  
	
	\textit{Acknowledgment}. This work was supported by the U.S. Department of Energy, Office of Science, Basic Energy Sciences, under Awards DE-SC0019068 and DE-SC0014208. The computational work was also supported by the Cypress High Performance Computing system at Tulane University, and by the National Energy Research Scientific Computing Center.
	\raggedright
	\bibliography{refs}
	
\end{document}


\title{Supplementary Materials: Thermodynamic metastability from weak interactions in the layered magnetic topological insulator MnBi$_2$Te$_4$}
\author{Jinliang Ning$^1$, Yanglin Zhu$^2$, Jamin Kidd$^1$, Yingdong Guan$^2$\\ Yu Wang$^2$, Zhiqiang Mao$^2$, and Jianwei Sun$^1$$^*$}
\date{%
	$^1$Department of Physics and Engineering Physics, Tulane University, New Orleans, Louisiana 70118, USA\\%
	$^2$Department of Physics, Penn State University, State College, Pennsylvania 16801, USA\\
	$^*$Corresponding author: jsun@tulane.edu \\[2ex]%
	\today
}
\maketitle

	\begin{figure*}	
		
		\begin{center}
			\begin{tabular}{ |p{2cm}|p{1cm}|p{2.25cm}|p{2.25cm}|p{1.5cm}|p{1.5cm}| }
				\hline
				\multicolumn{6}{|c|}{(a) Binding energies} \\
				\hline
				System & SOC & $E_b$(bulk) & $E_b$(mono) & $E_b/$f.u. & $E_b/$atom\\
				\hline
				Bi$_2$Te$_3$ & Yes & -812.56942913 & -270.70138786 & -0.1551 & -0.0310\\
				Bi$_2$Te$_3$ & No & -806.60272383 & -268.55431335 & -0.3133 & -0.0627\\
				MnBi$_2$Te$_4$ & Yes & -655.08006602 & -327.41573092 & -0.1243 & -0.0207\\
				MnBi$_2$Te$_4$ & No & -650.17902573 & -324.79699014 & -0.2925 & -0.0488\\
				MnTe & Yes & -113.50933 & -113.50933 & &\\
				MnTe & No & -112.62289 & -112.62289 & &\\
				\hline
				
			\end{tabular}
		\end{center}		
		
		\begin{center}
			\begin{tabular}{ |p{3cm}|p{3cm}|p{3cm}| }
				\hline
				\multicolumn{3}{|c|}{(b) Energy difference} \\
				\hline
				System & dE(bulk) & dE(mono) \\
				\hline
				Bi$_2$Te$_3$ & -0.3978 & -0.4294 \\
				MnBi$_2$Te$_4$ & -0.4084 & -0.4365\\
				MnTe & -0.4432 & -0.4432\\
				\hline
				\hline
				With SOC & $\Delta E=0.0711$ & \\
				Without SOC & $\Delta E=0.0895$ & \\
				\hline
			\end{tabular}
		\end{center}		
		
		\begin{center}
			\begin{tabular}{ |p{1.5cm}|p{1.5cm}|p{1.5cm}|p{1.5cm}|p{1.5cm}| }
				\hline
				\multicolumn{5}{|c|}{(c) Magnetic moments} \\
				\hline
				Form & SOC & Mn & Bi & Te\\
				\hline
				Bulk & Yes & 4.321 & 0.040 & 0.054\\
				Bulk & No & 4.330 & 0.043 & 0.048\\
				Mono. & Yes & 4.327 & 0.046 & 0.067\\
				Mono. & No & 4.337 & 0.052 & 0.052\\
				\hline
			\end{tabular}
		\end{center}
		\caption{Relationship between SOC and magnetic moment. (a) Binding energies in eV for bulk compared to monolayer, given per f.u. and per heavy atom. (b) Energy differences, per heavy atom: dE is the difference between total energies with and without SOC, while $\Delta E$ is the 0 K reaction energy. (c) Local magnetic moments for bulk and monolayer, with and without SOC.}
	\end{figure*}

	\begin{figure*}
		\centering
		\includegraphics[scale=0.7]{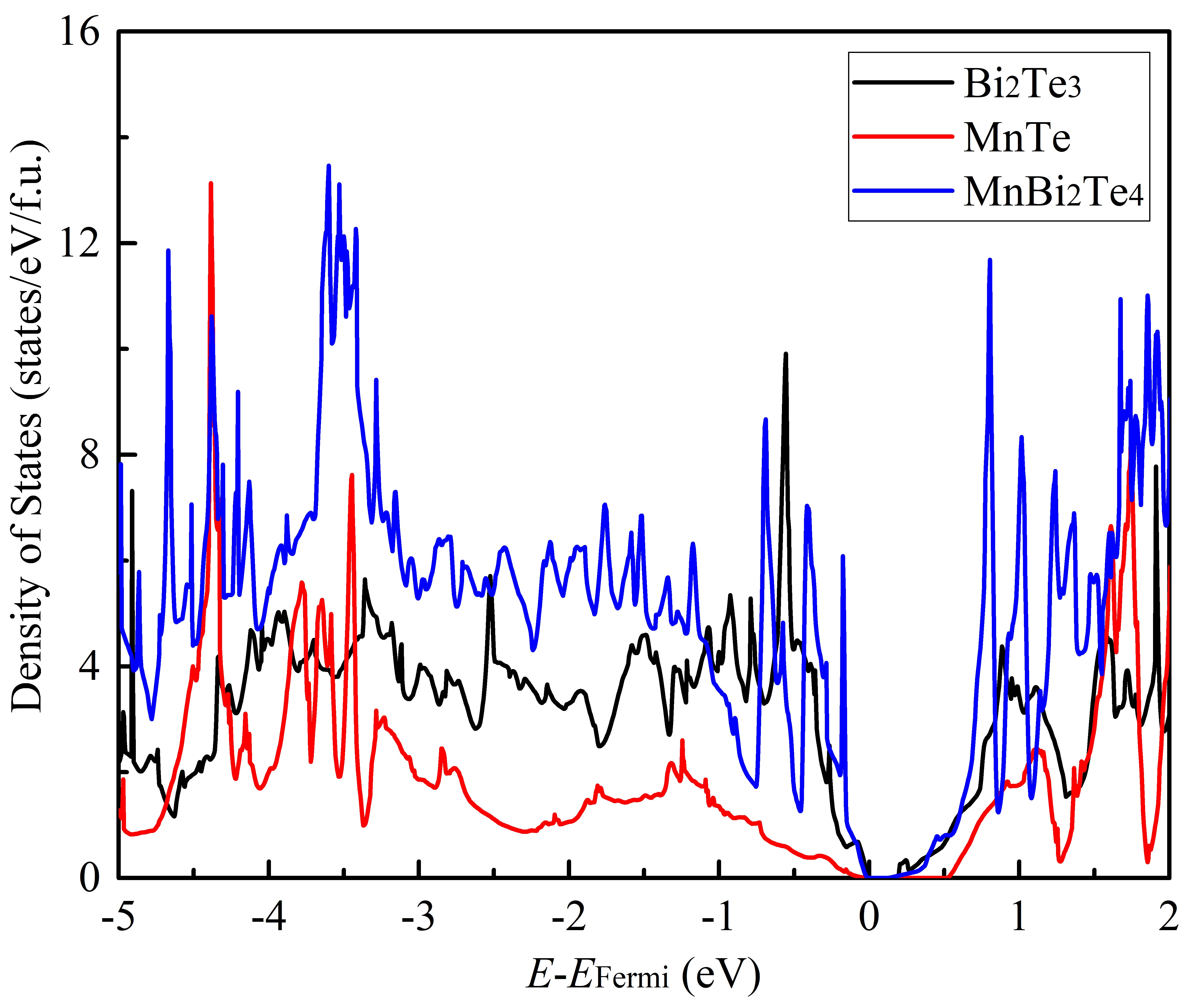}
		\caption{Electronic density of states for each phase, obtained from SCAN+rVV10 calculations with SOC.}
	\end{figure*}
	
	\begin{figure*}
		\centering
		\includegraphics[scale=0.7]{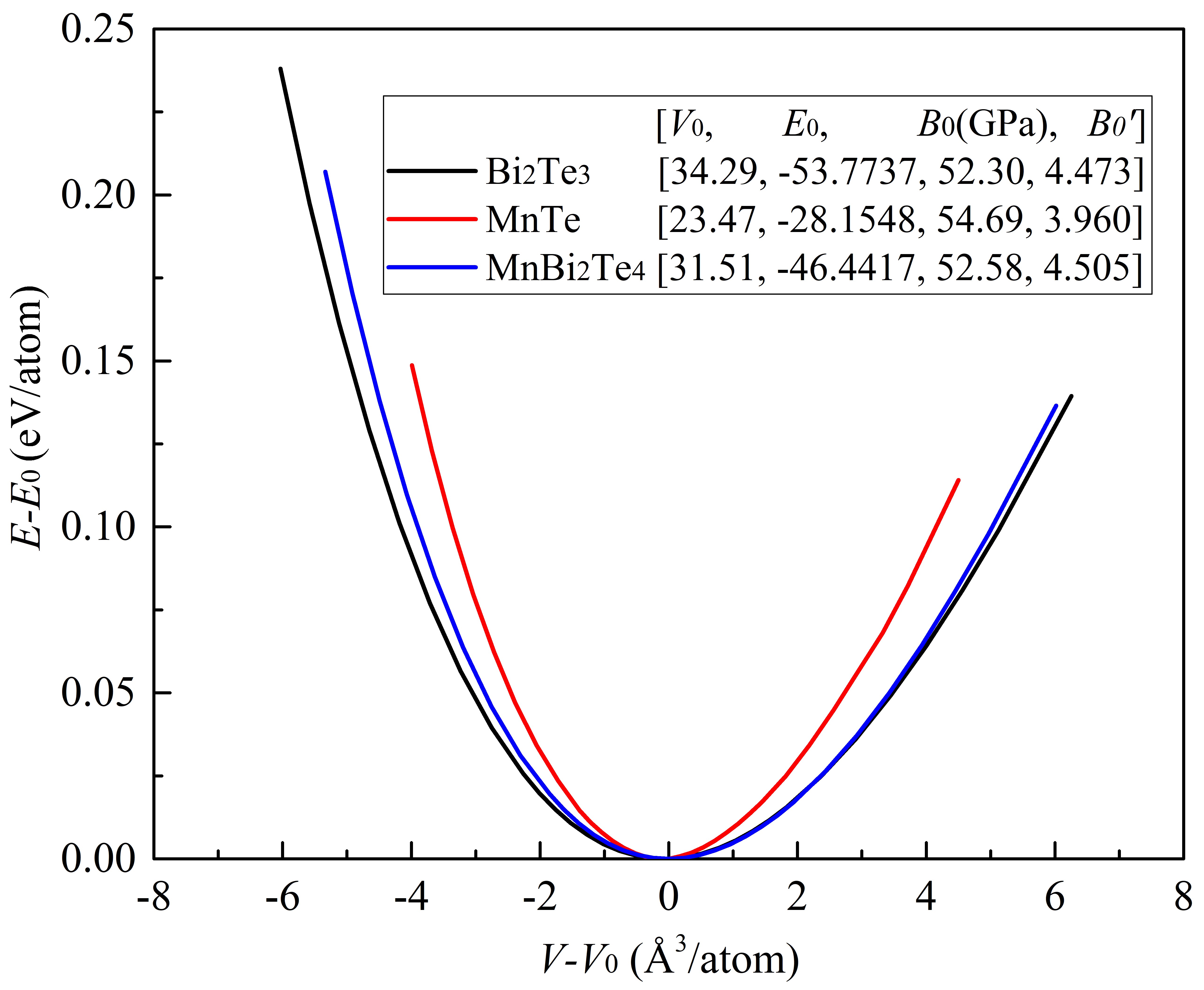}
		\caption{Static $E(V)$ curves for Bi$_2$Te$_3$, MnTe, and MnBi$_2$Te$_4$, plotted from SCAN+rVV10 results; (inset) fitting parameters for the Vinet equation of state in Gibbs2.}
	\end{figure*}
	
	\begin{figure*}
		\centering
		\includegraphics[scale=0.7]{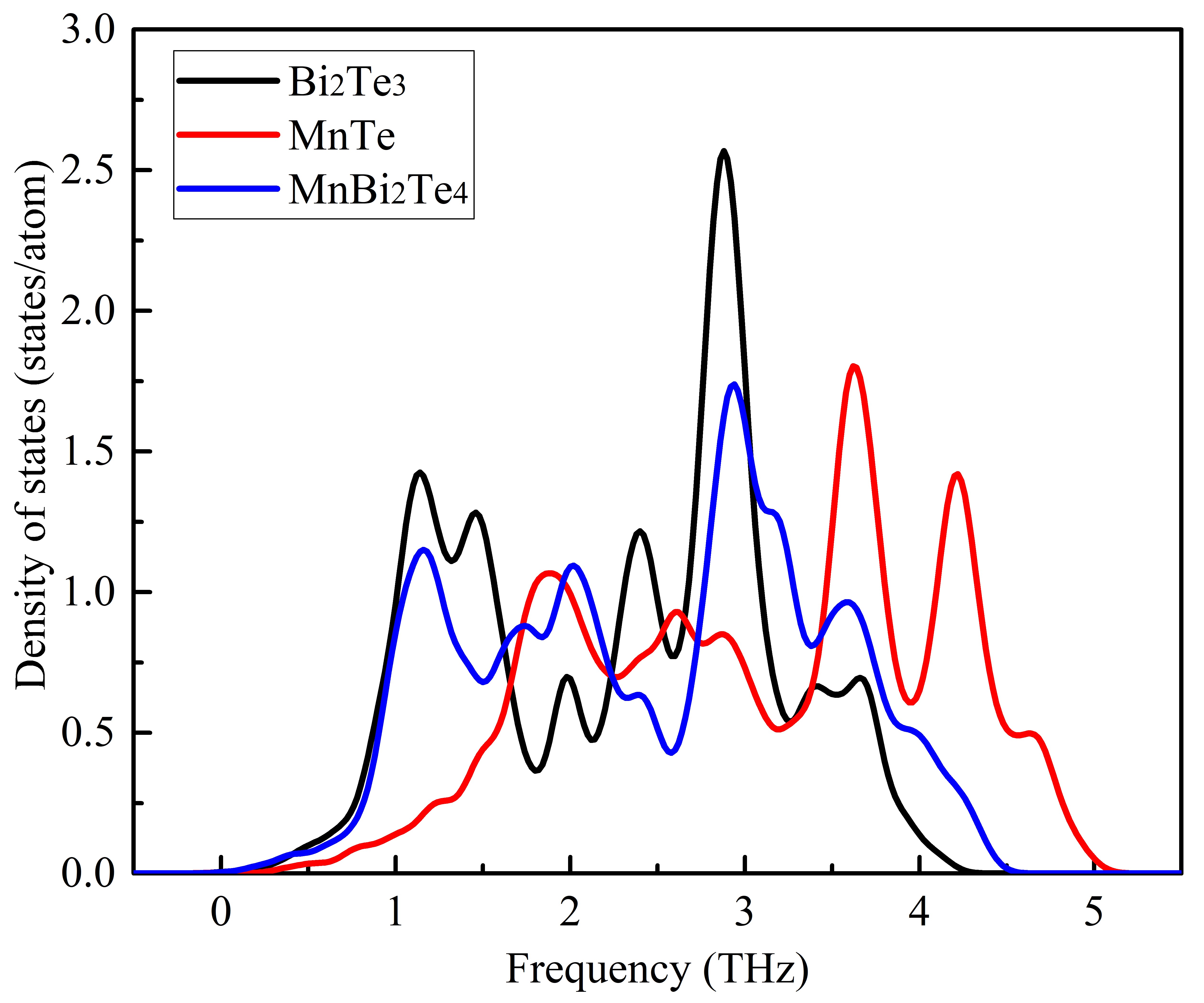}
		\caption{Phonon density of states for Bi$_2$Te$_3$, MnTe, and MnBi$_2$Te$_4$, obtained from PBE calculations.}
	\end{figure*}

	\begin{figure*}
		\centering
		\includegraphics[scale=0.65]{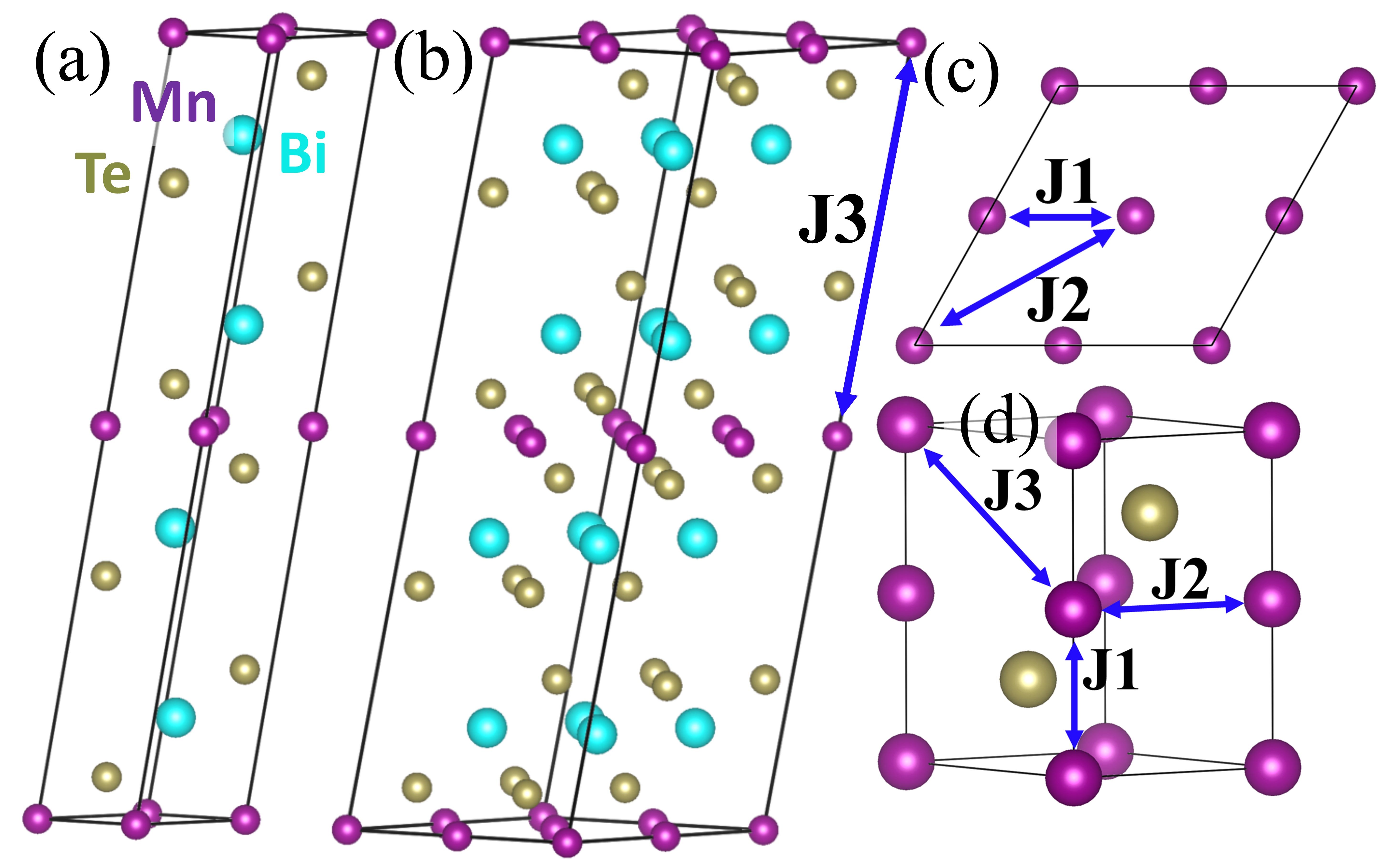}
		\caption{Crystal structure and magnetic exchange coefficients. (a) Conventional unit cell for antiferromagnetic MnBi$_2$Te$_4$. (b) $2 \times 2 \times 1$ supercell used in the $J_i$ calculations for MnBi$_2$Te$_4$. $J_3$ is the nearest neighbor (NN) interlayer exchange coefficient. (c) The NN and next-NN intralayer exchange coefficients $J_1$ and $J_2$ for MnBi$_2$Te$_4$. (d) The NN, next NN, and third NN exchange coefficients $J_1$, $J_2$, and $J_3$ for MnTe.}
	\end{figure*}
	
	\begin{figure*}
		
	\begin{center}
	\begin{tabular}{ |p{2cm}||p{2cm}|p{2cm}|p{2cm}|p{2cm}|p{2cm}| }
		\hline
		\multicolumn{6}{|c|}{(a) MnTe} \\
		\hline
		Structure & $E_{total}$ & $C_{J_1}$ & $C_{J_2}$ & $C_{J_3}$ & $C_{E_0}$\\
		\hline
		A & -450.4746916 & 8 & -24 & 48 & 1\\
		UUD & -337.7074274 & 6 & 6 & -36 & 0.75\\
		G-AFM & -450.2995318 & 8 & 8 & -16 & 1\\
		G-FM & -449.6535974 & -8 & 8 & 16 & 1\\
		\hline
		$J_{ij}$ (meV) & {} & -6.689 & 0.646 & -0.115 & {}\\
		$J_{ij}$ (K) & {} & -77.3 & 7.54 & -1.33 & {}\\
		Ref. (K) & {} & -21.5 & 2.55 & -9 & {}\\
		\hline
	\end{tabular}
	\end{center}

	\begin{center}
	\begin{tabular}{ |p{1.5cm}||p{2cm}|p{2cm}|p{2cm}|p{2cm}|p{2cm}| }
		\hline
		\multicolumn{6}{|c|}{(b) MnBi$_2$Te$_4$} \\
		\hline
		Structure & $E_{total}$ & $C_{J_1}$ & $C_{J_2}$ & $C_{J_3}$ & $C_{E_0}$\\
		\hline
		A & -2600.699990 & -24 & -24 & 8 & 1\\
		UUD & -1950.578785 & 6 & -18 & 6 & 0.75\\
		G-AFM & -2600.738875 & 8 & 8 & 8 & 1\\
		G-FM & -2600.742898 & 8 & 8 & -8 & 1\\
		\hline
		$J_{ij}$(meV) & {} & -0.359 & 0.164 & 0.040 & {}\\
		\hline
	\end{tabular}
	\end{center}

	\caption{First-principles analysis of magnetic coupling for MnTe and MnBi$_2$Te$_4$ using the N+1 method (Dalton Trans., 2013, \textbf{42}, 823-853). Unless otherwise indicated, energies are given in eV. Each $J_i$ is labeled in Figure S5. (a) N+1 method for MnTe. A $2 \times 2 \times 1$ supercell is used for A, G-AFM, and G-FM configurations, and a $\sqrt{3} \times \sqrt{3} \times \sqrt{3}$ supercell is used for the UUD configuration.  Our results are compared to previous CMC calculations (Phys. Rev. B \textbf{85}, 184413 (2012)). (b) N+1 method for MnBi$_2$Te$_4$. A $2 \times 2 \times 1$ supercell is used for A, G-AFM, and G-FM configurations, and a $\sqrt{3} \times \sqrt{3} \times \sqrt{3}$ supercell is used for the UUD configuration. For the N+1 method, the energies are determined as $E_{total}=\left(\sum_{i}C_{J_i}J_i\right)S^2 + C_{E_0}E_0,$ where $S=\frac{5}{2}$ and $C_{J_i}=-N_{J_i}\frac{\mathbf{S}_i \cdot \mathbf{S}_j}{S^2}.$}
	\end{figure*}